\documentclass[acmsmall]{acmart}

\AtBeginDocument{%
  }

\setcopyright{cc}
\setcctype{by}
\acmJournal{PACMHCI}
\acmYear{2025} \acmVolume{9} \acmNumber{7} \acmArticle{CSCW228} \acmMonth{11} \acmPrice{}\acmDOI{10.1145/3757409}

\usepackage[utf8]{inputenx}
\usepackage{color,soulutf8}

\begin{document}

\title[Fulfillment of the Work Games]{Fulfillment of the Work Games: Warehouse Workers’ Experiences with Algorithmic Management
}



\author{EunJeong Cheon}
\email{echeon@syr.edu}
\orcid{0000-0002-0515-6675}
\affiliation{%
  \institution{Syracuse University}
  \city{Syracuse}
  \state{New York}
  \country{USA}
}

\author{Ingrid Erickson}
\email{imericks@syr.edu}
\orcid{0000-0001-9841-8680}
\affiliation{%
  \institution{Syracuse University}
  \city{Syracuse}
  \state{New York}
  \country{USA}
}
  
\renewcommand{\shortauthors}{EunJeong Cheon and Ingrid Erickson}

\begin{abstract}
The introduction of algorithms into a large number of industries has already restructured the landscape of work and threatens to continue. While a growing body of CSCW research centered on the future of work has begun to document these shifts, relatively little is known about workers' experiences beyond those of platform-mediated gig workers. In this paper, we turn to a traditional work sector, Amazon fulfillment centers (FC), to deepen our field's empirical examination of algorithmic management. Drawing on two years of ethnographic research, we show how FC workers react to managers’ interventions, imposed productivity rates, and quantified objectification when subjected to labor-tracking systems in their physical work environments. Situating FC workers’ resistance to algorithmic systems and metrics within the current CSCW literature allows us to explicate and link the nuanced practices of FC workers to the larger discourse of algorithmic control mechanisms. In addition, we show how FC workers’ resistance practices are emblematic of `work games'---a long-studied means by which workers agentically configure (``trick'') their engagement within work systems. We argue that gaining a more nuanced understanding of workers’ resistance and consent in relation to algorithmic management expands our ability to critique and potentially disassemble the economic and political forces at the root of these sociotechnical labor systems.

\end{abstract}



\received{January 2024}
\received[revised]{October 2024}
\received[accepted]{February 2025}

\maketitle

\section{Introduction}

Recent CSCW scholarship has begun to document the significant shift in work wrought by the proliferation of technologies like algorithms~\cite{wolf2019evaluating,lampinen2022cscw}, robots~\cite{cheon2022working, cheon2022robots}, and digital platforms~\cite{qadri2021s,jarrahi2020platformic}. This shift has intensified the datafication of workplaces, from office settings with employee monitoring software and productivity analytics~\cite{dencik:2023regimes} to advanced “smart warehouses”~\cite{sanchez:2019datafication}. The rise of data-centric and automated enterprise systems is influencing both organizational dynamics and individual worker experiences~\cite{lee2019procedural,wang2019didi,khovanskaya2019tools} across various industries. Central to this transformation is the concept of algorithmic management, which has become a focal point of scrutiny and critique within the field~\cite{kumar2018uber,gregory2021my,de2021lose}.  To date, algorithmic management has primarily been studied within the context of labor platforms, such as Uber or Upwork. Less research has focused on issues of datafication and control within traditional workplaces, such as warehouses~\cite{cheon2024amazon, cheon2024examining}. We suggest that the unique amalgamation of datafied physical spaces and surveilled digital environments represents an important, missing case for investigating the impacts of algorithmic management and the future of work~\cite{cheon2023powerful,cheon2021human}. In these settings, such as Amazon Fulfillment Centers (FC), workers navigate both a highly controlled physical world and a virtual one, subject to monitoring by managers and data systems alike. These environments are not just sites of labor but also of intricate interactions between white-collar managers and blue-collar workers, each group sharing a workspace that is governed by both tangible and virtual controls and engaging with technology and data in distinct ways~\cite{cheon2023powerful,cheon2024amazon}. This paper delves into the experiences of workers within these algorithmically managed FC, focusing on their interactions with algorithmic management systems and their responses, including resistance, to these pervasive data-driven metrics and controls. By examining FC as a microcosm of the increasingly datafied workplace, we gain insights into the broader implications of technology-driven management structures on worker experiences and organizational practices.

Amazon fulfillment centers (FC)\footnote{For clarity, the terms “FCs” and “Amazon warehouses” are used interchangeably throughout the paper.} have long been some of the most technologically rich workplaces in existence~\cite{BBCNews2020,day_2022,amazon-ML-warehouse_2019}, where advanced algorithms and robots are brought in for testing and deployment. For example, Amazon placed itself at the forefront of a robotics race by deploying more than 520,000 robotic drive units across its fulfillment and sort centers over the last decade. This move is expected to alter the future of the warehouse sector fundamentally. In parallel, Amazon FCs' in-house automated system, in which algorithmic technologies direct the work of thousands of human workers, offers a glimpse into how traditional workplaces such as warehouses may soon evolve~\cite{amazon-robot_2017}. Further, Amazon's strategic amalgamation of technologies into sets, including cameras, scanners and sensors, creates a template for the monitoring, collecting, and measuring of worker productivity and the active prevention of worker misbehavior. As a mechanism to ensure that Amazon can deliver on its ever-increasing promises to Prime members about same-day and next-day package delivery, automation and AI technologies are, at their very core, about maximizing the company’s profits. 

Notably, Amazon's warehouse practices are not occurring in a vacuum. With its enormous market share and research-and-development power, Amazon increasingly sets the standards for algorithmic productivity and efficiency that other companies seek to replicate. Many e-commerce and logistics companies take pride in “Amazonifying”~\cite{kang_2022amazonification} their business operations to emulate Amazon's perceived advancements. These reputation-based effects set the stage for investigating the experiences and challenges that FC workers currently face as Amazon's algorithmic controls will likely define and legitimate nearly all datafied warehouses and worker experiences in the near future.


This paper is based on two years of ethnographic research conducted with workers at Amazon fulfillment centers~\cite{cheon2024amazon,cheon2024examining}. In addition to observing FC workers' online communities, we conducted 16 interviews with FC workers. In our findings, we show how Amazon's labor-tracking system involves the interplay of productivity metrics, various devices, and barcodes on products and workers’ badges. Furthermore, we describe the discrepancies between the labor tracked by the system and the actual labor input, which require human intervention to resolve. The findings also reveal how the interplay and discrepancies tend to reify human workers and their labor as numbers. Lastly, we discuss how workers strategically resist the labor tracking system in the form of ``tricks''---a phrase in their own vernacular\footnote{Some concepts with a similar meaning but a more neutral connotation are present in the CSCW literature, such as ``coping strategies''~\cite{jhaver2018algorithmic}. Following the principle of polyvocality in ethnographic writing, we use the actual words of the workers instead of paraphrasing or rephrasing from our own analytical point of view.}.



This paper makes several important contributions. First, we provide a detailed description of warehouse work as organized by big tech. We highlight this workplace as one that is increasingly equipped with algorithmic technologies and metrics in order to execute both old and new forms of management and control. Our ethnographic research illuminates how FC workers react to labor-tracking system and metrics, deal with managers' interventions, acclimate to imposed productivity rates, and digest the company's treatment of them as numbers. With this detailed account of FC workers, we extend the scope and import of CSCW research on workers' experiences under algorithmic management. While extant CSCW research provides a rich understanding of platform and digital labor (e.g., ride-hailing platform workers~\cite{watkins2020took,sehrawat2021everyday}, food-delivery workers~\cite{kusk2022platform}, online freelancers~\cite{wilkins2022gigified}), the entanglement of conventional work settings\footnote{The term ``conventional (or traditional) work sector'' in this paper encompasses industries where work organization does not primarily rely on digital platforms~\cite{jarrahi2021algorithmic}. Traditional coordination mechanisms, such as organizational hierarchies and ``direct employment'' (p. 577) and a stronger temporal connection to the organization~\cite{cappelli2013classifying}, are characteristics of the traditional work sector. Thus, by having at least one person who is directly responsible for monitoring, evaluating, and motivating their performance, the employees work within the organization's control system, aligning their ideas about reward, punishment, and evaluation with those of the organization.}  and algorithmic management has been too often disregarded to date. This paper addresses this omission. Of equal importance, we bring our findings into conversation with existing literature (both within CSCW and from adjacent fields) where algorithmic management has been theorized primarily through the prism of platform labor. Reversing the move that sociologist Wood~\cite{wood2021:algorithmic} took when comparing detailed case studies of platform work with surveys of conventional work, we delve into conventional work first as a elucidating mechanism for investigating the implications of policy and working conditions for the future algorithmic workplace.






Finally, we raise the construct of ``work games'' to the fore as a lens for understanding the deeper meanings of resistance in response to algorithmic management. Sociology and organizational studies have long viewed work(place) games as mechanisms that workers employ to create the illusion that they have real control over their daily activities under capitalism~\cite{burawoy1979manufacturing}. 
Sociologist Michael Burawoy conceptualized the term ~\textit{work games} based on his ethnography of an aeronautics factory. He showed how workers create ``games'' to pass the time and endure tedious, meaningless work. He observed that workers engaged in  ``making out,'' which involved an improvisational game invented to maximize earnings by speeding up machinery in order to surpass quotas. In his analysis, Burawoy highlighted the fact that neither coercion nor additional compensation could adequately explain the dedication of these factory workers to the game of ``making out.'' Workers at his field sites, including himself, were preoccupied with ``winning'' the game. He attributed this to the fact that winning provided workers with psychological and emotional satisfaction (e.g., helped them counter the frustration that arises from repetitive work). And yet, Burawoy also noted that the game of ``making out'' also led workers to align with management in the production of greater surplus value for the company. As such, the concept of a `work game' encapsulates the complex reality that many workers exist in---games are at once a relief as well as a means to participate in the management's goals. Paradoxical as it may sound, Burawoy discovered that ``making out'' is not the result of either capitulation or coercion; both of these reasons fail to explain the inception and persistence of this phenomenon in reality. Instead, making out has dual affordances: one that benefits the worker and the other that benefits the employer.
We draw on the construct of the work game (e.g.,~\cite{burawoy1979manufacturing,ranganathan2020numbers,mccabe2020changing}) to analyze FC workers' resistance to the labor-tracking system in a new way.  We also show how workers' individual ``tricks'' configure their engagement (and consent) within the system in a different way than has been analyzed before. Applying this lens helps us to elucidate three forms of worker resistance: active consent-resistance, passive consent-resistance, and non-consenting resistance. These concepts build a new analytical foundation that aligns with the discursive relationships between workers' consent and resistance found in our data. 
We argue that establishing how workers' resistance is expressed as consent explains workers' experiences and reactions to algorithmic management with greater nuance and accuracy, at least in the context of traditional workplaces.


\section{Related Work: Algorithmic Management and Control in Data-Driven Warehousing}

Over the last decade, scholars in CSCW and related fields have begun to publish a growing body of literature related to algorithmic management and algorithmic control. These scholars define algorithmic management as “the delegation of managerial functions to algorithms”~\cite{lee2015working,lee2018understanding,noponen2019impact} and refer to algorithmic control as the automation of managerial practices~\cite{tomprou2022employment} or the ability of algorithms to supervise workers’ performance in alignment with an organization’s goal~\cite{mohlmann2021algorithmic}. While the two terms are often used interchangeably in prior literature (e.g.,~\cite{barati2022effects,kellogg2020algorithms,zhang2022algorithmic}), ``algorithmic control'' generally tends to refer to specific managerial functions, such as the act of monitoring and controlling workers. It has also been described~\cite{pregenzer2021algorithms,adensamer2021computer} as the use of algorithms to guide and control workers in a way that human managers would. Much of the scholarship on algorithmic control draws on Edwards’s concept~\cite{edwards1978social} of managerial control---i.e., managers’ attempts to control and manage any conflicts of interest with workers. In this way, the term implies that algorithmic technologies can afford managers new mechanisms to exert their control, such as in the case of utilizing extensive electronic surveillance. In addition to controlling workers, algorithmic technologies can also be used to discipline, evaluate, and direct employees. This paper adopts the term ``algorithmic management'' rather than the more specific term ``algorithmic control'' in order to encompass all of these roles.


\subsection{Worker Experiences with Algorithmic Management and Control}

While recent CSCW and HCI studies offer a rich understanding of workers’ experiences under algorithmic management, they draw primarily on platform-mediated work or gig work (e.g., Uber~\cite{kumar2018uber,haque2021exploring}, Upwork~\cite{munoz2022platform,hulikal2022collaboration}, Deliveroo~\cite{kusk2022platform}, Amazon Mturk~\cite{gray2016crowd,irani2013turkopticon}) to develop their understanding of this phenomenon. For example, research has been done on workers’ interpretations of biometric security integrated into ride-hailing platforms~\cite{watkins2020took}, gig delivery workers’ perceptions of their relationships with different stakeholders mediated by algorithms~\cite{chen2022mixed}, platform freelance workers’ experiences of a reputation system-based algorithmic management~\cite{holtz2022much} and Uber drivers’ everyday interactions with the Uber platform in the Global South~\cite{sehrawat2021everyday}. These studies shed light on various challenges and struggles that workers experience in terms of their well-being~\cite{zhang2022algorithmic}, emotion labor~\cite{raval2016standing}, social isolation~\cite{seetharaman2021delivery}, limited career mobility~\cite{blaising2022managing}, gender-related bias and harassment~\cite{ma2022brush}, and the feminization of labor as it is reinscribed by platforms~\cite{huber2022approximation}. While much of this work tends to take a negative view of algorithmic management, some studies note that algorithmic control and surveillance can also benefit workers in some ways, such as providing an increased sense of safety for women gig workers~\cite{anjali2021watched}. As our collective understanding of algorithmic management has developed over the years, its articulation has become more nuanced and is increasingly expressed as occurring along a spectrum. Kusk and Bossen~\cite{kusk2022working}, for example, describe a softer form of algorithmic management, which they dub “lenient algorithmic management.”


 The role of algorithms in conventional work sectors such as manufacturing, retail, hospitals~\cite{parent2022algorithms}, marketing firms, call centers~\cite{brione2020my,call-center_2018}, big tech warehouses~\cite{delfanti2021machinic,gent2018politics}, hotels~\cite{orlikowski2014happens}, and major banking firms~\cite{keller2017influence} has also begun to be documented in recent years. A set of rich ethnographic studies detailed the rise of algorithms in newsrooms~\cite{christin2017algorithms} and criminal justice institutions~\cite{brayne2021technologies}, with a specific focus on how these organizations receive and incorporate algorithmic tools. This research shows that work settings outside of platforms typically involve a direct, often physically-proximate, relationship between employer and employee~\cite{cappelli2013classifying}. They also tend to be structured more variably with regard to organizational hierarchy than platform-based work contexts.


In non-platform settings, most algorithmic management systems are designed and utilized to support employee decision-making. In this way, they can be seen as evolutionary advancements of the decision support tools of earlier decades. For example, the predictive algorithm tools used in modern courtrooms today ~\cite{brayne2021technologies} are used to provide calculated forecasts of violent recidivism risk. In the context of journalism, web-based analytics ~\cite{christin2017algorithms} offer writers and publishers real-time data about visitors' activities for each article. Similarly, algorithmic systems are used to assign tasks to employees, as in the case of flexible scheduling tools that create tasks for retail employees ~\cite{oort_2018} based on customer demand and required skills. In these examples, algorithmic tools do not directly target individual productivity or performance, nor do they collect and report data on employees during their work. There are a few exceptions, however, including automated performance feedback systems in call centers~\cite{brione2020my, call-center_2018}, which increasingly guide agents in real-time to improve their voice tones and phone etiquette skills.

Among the various non-platform work settings mentioned above, warehouses might vie with call centers as the locales in which algorithmic monitoring and performance management reign most supreme~\cite{wood2021:algorithmic}. In warehouses, algorithms are used regularly to generate employees' productivity rates and real-time work pace. As previous warehouse studies  show~\cite{bloodworth2018hired,mcclelland2012warehouse,moore2016quantified}, worker performance is primarily evaluated according to a set of predetermined productivity metrics. Workers are required to perform at a pace stipulated by these metrics, which are then measured against the data captured by a worker's handheld device. Algorithms are used to score workers and these scores are then used by managers as a basis for sanction or termination~\cite{briken2018fulfilling}. In Amazon fulfillment centers, the focus of our study, we consider these algorithmic technologies for monitoring and controlling workers as labor-tracking systems and metrics.

Given Amazon's outsized role in the global economy, algorithmic management in Amazon's warehouses, in particular, has garnered attention from scholars in the fields of sociology, labor studies, and management. These studies provide valuable insights regarding logistic media~\cite{beverungen:2021amazon}, labor control strategies~\cite{fuchs2022location,delfanti2021machinic}, working conditions in warehouses~\cite{schein2017taylorism,briken2018fulfilling}, and workers' participation in strikes~\cite{apicella2019divided,cattero2018organizing}. For instance, Schein~\cite{schein2017taylorism} conducted a theoretical examination of how Amazon incorporates the principles of Taylorism, resulting in the dehumanization of its workforce. Briken and Taylor~\cite{briken2018fulfilling} examined the experience of warehouse workers in light of job-scarce labor markets, the involvement of temporary worker agencies, and the impact of workfare and benefit sanctions. Sociologists Vallas and Kronberg~\cite{vallas:2023coercion} investigated Amazon’s labor practices and their impact on workers' perceptions of their jobs and class status, and in related scholarship~\cite{vallas:2022prime} Vallas and colleagues theorize how both coercive and hegemonic labor control mechanisms are tailored to target employees in varying precarious positions. These studies reveal that the application of algorithmic technology is particularly effective at monitoring workers unaccustomed to stable employment and lacking alternative job options. It also amplifies the increasing potency of algorithmic control across myriad work environments---a phenomenon these authors refer to as “techno-economic despotism”~\cite{vallas:2022prime}. 

Put together, this research also suggests that physical work environments, such as Amazon warehouses, are defined by a clear hierarchical order. This characteristic renders these locales particularly salient for understanding workers' struggles for rights in relation to algorithmic management. While empirical, in-depth social scientific analysis in this area is on the rise~\cite{jarrahi2021algorithmic}, the overall discussion of algorithmic management's effects remains overshadowed by hearsay and journalistic reports~\cite{vallas:2022prime}. More broadly, few existing algorithmic management studies examine the lived experiences of workers. Highlighting this perspective is paramount for gaining a fundamental understanding of how individuals interact with and perceive these systems~\cite{cheon2024examining}. 

In this paper, we take up the call to investigate workers’ experiences with algorithmic management by centering our analytic gaze on the datafied warehouse environments of Amazon FCs. Incorporating workers' personal narratives, we present an ethnographic study that methodologically captures the firsthand experiences of workers---a perspective that is often underrepresented in large workplace studies. Such a focus enables us to document precisely what workers see, hear, and do daily. Building upon this empirical base, we then draw meaningful comparisons across different contexts and periods in our data to identify the sociotechnical dynamics at the intersection of algorithms, human managers, and worker agency within the FC. In so doing, this study both broadens and deepens the current understanding of algorithmic management by showing 
that algorithmic tools, despite the degree of their designed application, are always situationally enacted.

\subsection{Worker Reactions and Resistance to Algorithmic Management and Control}

Although many studies on algorithmic management focus on the influence it has on workers and work arrangements, a new cache of research is beginning to redirect attention to the ways in which workers react in algorithmic work environments~\cite{rahman2021invisible,jarrahi2019algorithmic,lee2015working,mohlmann2017hands}. 
For example, a set of recent CSCW studies showcase how workers, both individually and as collectives, develop strategies to address conditions of algorithmic management. Looking at the experiences of workers on Upwork, Jarrahi and Sutherland~\cite{jarrahi2019algorithmic} elucidate three distinct strategies: sensemaking, circumventing, and manipulating platform algorithms. The authors show how workers develop “algorithmic competencies” to deal with and appropriate algorithmic management---a fact they assert demonstrates that platform workers are not just “passive recipients” of algorithmic control. Kusk and Bossen’s research~\cite{kusk2022working} on food delivery workers describes the optimization strategies that couriers use, including how they optimize their assigned tasks, working hours, and vehicles. The study further shows that local support structures (e.g., workers’ social networks) are vital in providing workers with the necessary tools and knowledge to engage in these optimization strategies. ~\citet{kinder2019gig} similarly highlight how freelancers at Upwork make use of “ecosystemic alliances,” namely external digital platforms that enable workers to both resist and protect themselves from algorithmic control and surveillance. Other studies have also identified the ways in which gig workers develop “workaround strategies”~\cite{lee2015working} to cope with and regain control under platform algorithms~\cite{sallaz2015permanent,jhaver2018algorithmic,eslami2017careful}. Digital economy scholars have built on this empirical research to create a three-part classification of strategic engagement with algorithms: manipulation, subversion, and disruption~\cite{ferrari2021fissures}.  

 
Scholars in sociology, STS, and labor studies have also been active conceptualizing various reactions and resistant tactics by workers under algorithmic management recently. For example,  ~\citet{valeria2022control} envision workers regaining control on labor platforms as a “space” when they can manipulate the algorithms to their advantage. Other research classifies workers' reactions as either emotional or actional~\cite{langer2021future,pregenzer2021algorithms}. Relatedly, the reactions laid out by ~\citet{pregenzer2021algorithms}---blending, bridging, distancing, and separating---for instance, show that workers sometimes need to cooperate with the algorithms by embracing controls (blending); they also fill gaps in relation to inscrutable controls by bridging. When describing workers gaining or regaining control, STS and labor studies researchers often use the metaphor of a game. Here, the act of gaming the system typically refers to workers manipulating the “loopholes in the systems” for their own benefit~\cite{mohlmann2017hands}. Regarding these loopholes, ~\citet{ferrari2021fissures} suggest the concept of “fissures in algorithmic power,” referring to “moments in which algorithms do not govern as intended” (p. 815). They describe platform workers exploiting these fissures in their favor through acts such as manipulation, subversion, and disruption. Other researchers use the term “resistance” to highlight workers’ ability to secure agency in the face of algorithmic management. Drawing insights from research on traditional service work sectors, ~\citet{cameron2022expanding}, for example, discusses workers’ “covert resistance” to a ride-hailing platform that is “hard-to-observe actions that undermine organizations’ attempts at control.” Relatedly, ~\citet{kellogg2020algorithms} label individual and collective resistance of algorithmic control as “algoactivism.” 
 
Despite this emerging scholarly attention to workers’ reactions and resistance on platforms~\cite{shapiro2018between,maffie2022perils,rahman2021invisible}, little is known about how workers develop their resistance tactics in physical work settings like warehouses. Compared with platform workers, datafied warehouse workers encounter the dual challenge of navigating on-site managers’ supervision as well as the real-time monitoring and evaluation deployed by omnipresent labor-tracking systems. Moreover, the various devices connected to the algorithm systems (e.g., handheld devices, internal apps) impact how workers “sense-make”~\cite{jarrahi2019algorithmic} in these sociotechnical contexts. Our close examination of FC workers aims to provide new insights into the broader conversation on algorithmic management that can be applied to other increasingly-datafied work settings, particularly those that combine both traditional and algorithmic forms of managerial oversight.

\section{Method}

This work is grounded in the first author’s two-year ethnographic research on Amazon FC workers. The ethnographic research consists of online participant observation, interviews, reviews of archival materials from FC workers’ unions (e.g., the unions' statements, newsletters), and a visit to a union event. The study broadly focuses on workers’ experiences with emerging technologies in the context of their working lives, specifically the effect of labor-tracking systems as management tools. 


We briefly describe the overall trajectory of the ethnographic research, which was conducted in the form of a multi-sited ethnography in an effort to “follow the people”~\cite{marcus1995ethnography}: the workers at Amazon FCs. The field sites for this investigation are three online communities organized and run by FC workers. These communities act as safe spaces where FC workers get together and freely share their experiences\footnote{While direct access to physical FCs was restricted, limiting engagement with labor-tracking systems and metrics, ethnographic research does not require physical presence in a bounded site to generate rich insights. As Gusterson’s \textit{polymorphous engagement} approach demonstrates~\cite{gusterson1997studying}, ethnographers have long relied on dispersed, networked sources—including digital spaces—to trace cultural practices and embodied experiences. This study builds on that tradition, centering workers’ own narratives of life inside FCs while addressing both digital and physical dimensions of their work. Moreover, securing access through a partnership with Amazon would have imposed constraints on research outcomes.}.


\subsection{Ethnographic Research}

\subsubsection{Online Observations}

The research began in September 2020 with online observations of the three online communities (two subReddits and one Discord channel). At the time, these communities were the only detectable ones that were open for the public to join. To preserve the privacy of these communities and their members, we do not identify them by name here. The primary purpose of observing each of these communities was to grasp what was “going on”~\cite{goffman1974frame} in the warehouses as articulated emically by FC workers. Secondarily, observations also allowed the first author to identify FC workers for interviews. Over the course of a year, until September 2021, the first author spent 2-3 hours a week reading the postings and comments posted by workers across the three communities. Most postings focused on work-specific questions or concerns. For example, posters would ask about work policies (e.g., voluntary time-off policy, vacation), workplace regulations (no headphone/cellphone rules), or seek advice on certain tasks (how to handle poorly stowed items). Other posts revealed details about the various technologies (e.g., the A to Z app, the hand scanner) used in the FCs; these were objects of frequent discussion and debate within each of the communities. To ground her understanding, she created a glossary of the FC terms and acronyms we encountered, such as TOT (time off tasks) or UPH (units per hour). she also captured interesting posts and replies as screenshots and saved these along with brief field notes. 
 
After establishing a baseline understanding of the discussions occurring within in the three FC communities, the first author moved on to engage in two periods of intensive non-participant observation~\cite{boellstorff2012ethnography}. The first period spanned September and October 2021 and the second lasted from May to June 2022. During these observations, the first author spent at least twenty hours a week in the online communities. During this time, she monitored the daily community discussion on each site by taking notes of interesting aspects, documenting questions, and noting recurrent themes. Furthermore, she extracted all postings and threads made over the course of a one-year period (from November 1, 2020 to October 31, 2021) from each site. The 2500 postings from each community with the most replies were also collected (1998 of them from distinct users).



\subsubsection{Interviews}

The intensive observations prepared us to engage in timely and salient interviews with FC workers. The first author recruited 16 participants from the three online communities, where, as a verified member, she received approval from the communities' moderators for the recruitment. In order to protect the anonymity of the interviewees, we report on the group of interviewees as a whole in lieu of using a table to describe each interviewee individually. All workers in the study started as level one associates (the lowest rank of position) and have worked as Packers (four interviewees), Stowers (six), or Pickers (six). Most interviewees (eleven of sixteen) have had more than one role (e.g., working both as a Packer and Problem Solver or both as a Stower and Learning Ambassador). Nobody holds the position of manager. Their working periods at FCs range from 6 months to 5 years (at the time of the interview), and they are employed at sixteen different FCs in twelve different US states. The age of the interviewees ranges from late teens to 30s, although most (thirteen) are in their 20s. 
The gender distribution of the interviewees is 4 female, 10 male, and 2 non-binary.
 
To construct a thick account~\cite{geertz1973interpretation} of the experiences and stories that each FC worker had to share, the first author conducted each interview following the Internet ethnographer Christine Hine’s~\cite{hine2020ethnography} “behind the screen” recommendation.  During the interviews, discussions ranged from workers sharing their experiences with different technologies in the FCs; articulating concerns or tensions surrounding the technologies, productivity metrics, regulations and policies; and imagining future opportunities for worker empowerment. The average interview lasted 81 minutes (minimum: 55 minutes, maximum: 107 minutes) and was conducted using a video conferencing tool. All interviews were recorded and transcribed for analysis purposes, and each interviewee was compensated for their time with a \$30 Visa gift card.


\subsection{Data Analysis and Researcher Positionality}

Our analysis followed a constructivist grounded theory process~\cite{charmaz2006constructing,charmaz2021genesis}, which was positioned from the standpoint of “empathetic understanding”~\cite[p.292]{charmaz2011constructivist}. We carefully read and analyzed all of the assembled data (i.e., observation notes, screenshots, images, interview transcripts) by engaging in two rounds of line-by-line coding, memoing, and constant comparative iteration between data and codes. Our initial codes focused on the processes and actions expressed by the FC workers ~\cite{charmaz2006constructing} (e.g., continuously carrying the app for tracking purposes, checking if the tracking system was recording any work inactivity). Our second round of codes involved the refinement and in-depth examination of the meanings intertwined with these workers' practices, such as the constraints imposed by labor-tracking metrics. This two-tiered coding approach honed our conceptual understanding of the data. 


\subsubsection{Methodological Self-consciousness}

It bears comment that our personal and research experiences, background, research orientations, and values all shaped our analytical approach and our ability to synthesize new insights~\cite{charmaz2017power}. The first author was educated in the US and primarily uses ethnographic methods as a CSCW and HCI researcher. She has prior experience studying social dynamics among workers in labor-intensive workplaces (e.g., manufacturing factories that utilize new technologies such as robots~\cite{cheon2022robots,cheon2022working}), but she has never had the lived experiences of being a blue-collar worker. She approaches the work presented in this paper through a constructivist-interpretive paradigm, with the primary goal of gaining a deeper understanding of the world of human experiences. For the first author, getting the workers’ stories and perspectives posed a rather challenging endeavor. Recruiting workers even for interviews frequently failed: many times, the individuals contacted did not believe that the author was not associated with Amazon or they simply declined to talk with her. Most mentioned their employee non-disclosure agreement with Amazon or that they had no energy to talk with the author, even when being told that they would be compensated for participating. She was ultimately able to connect with the workers for scholarly work and compensate them adequately for their time sharing their stories and perspectives.

The second author is also an ethnographer, who comes to this research by drawing on her experiences studying how workers often hack and circumvent the sociotechnical situations they find themselves in. She also has no common experience with FC workers as she has never worked in a warehouse nor been subject to algorithmic forms of control in her work. Like her co-author, she sees an opportunity with this research to expand the discourse on algorithmic management beyond the work contexts that have become overly familiar to CSCW researchers. 

As previously noted within the CSCW field, conducting ethical research necessitates being aware of and responding to a multitude of small details during the research process in order to identify ``ethical encounters''~\cite{waycott:2017ethical}. Some readers may be concerned about the potential repercussions of reporting our workers' tactics (or ``tricks'') in response to the labor-tracking systems. For instance, the company could modify its systems to prevent the tricks. This is not outside the realm of possibility. We expect, however, that the workers in our study may not have divulged some tricks that could be identified and corrected by the company in the future. Moreover, as detailed in the section that follows, the shared tricks by the workers are largely in line with the management's interests (i.e., maintaining high productivity rates). In this regard, it is unlikely that the company will take action. Nonetheless, we can imagine the worst-case scenario being that current tricks will no longer work in the future necessitating the need for workers to create new ones.

\section{Findings}

To set the scene for a discussion of our findings, it is necessary to first provide a detailed overview of the productivity metrics and labor-tracking system implemented in FCs. The work productivity of an individual FC worker is calculated by the number of tasks completed per hour, which is called \textit{rate} by the FC workers and managers. The workers are expected to ‘make rate’ of a certain number; otherwise, they may face disciplinary action, up to and including termination of employment. When a worker is tasked with the responsibility of scanning an item, their productivity is measured by their ‘scan rate,’ which denotes the number of items successfully scanned by them within an hour. ‘Stowing’ is the process of loading items onto pods composed of stacked totes and bins while ‘picking’ is the process of retrieving items from pods. The metrics used to measure the efficiency of these tasks are referred to as the ‘stow rate’ and the ‘pick rate’ respectively. We learned from FC workers’ online discussions that a rate of over 200 (units per hour for picking) is the standard when one first gets hired, while a rate of 420 is expected during peak seasons. The online community discussions frequently centered on questions and descriptions of struggles related to making the rate, such as individuals seeking tips on how to increase their rate and expressing frustration about falling behind. 

The rates based on productivity metrics are measured, monitored, and presented with the help of various technologies surrounding FC workers. In this paper, we refer to the assembled collection of technologies deployed for tracking individual workers’ rates as the \textit{labor-tracking system}, despite each individual technology serving a distinct function. Specifically, the labor tracking system in an Amazon FC includes the following components: individual barcodes; handheld scanners that track workers’ activities, speeds, and locations; workstation computers with software developed to evaluate work performance; monitors at workstations constantly displaying individual workers’ performance; cameras placed strategically throughout a warehouse to monitor worker movements; the “A-to-Z” application where workers receive notifications on their performances or messages from HR or the management; and clock-in/clock-out stations. Managers can also track each employee’s progress and productivity level in real time using software on their laptop such as the Associate Development and Performance Tracker.

In the following section, we illuminate the ways that Amazon's labor-tracking system is a dynamic interplay of productivity metrics, devices, and barcodes that act upon (almost all) objects in the FC, including  products, product containers, shelves, and workers' badges. We also describe what happens when there is a discrepancy between the activities tracked by the system and the actual labor input---a friction point that requires human involvement to resolve. Our findings further reveal that the dynamics constituted by and surrounding these systems obscure human workers and their labor as numbers. Lastly, we share what kinds of tricks human workers utilize in response to being numerified and, more broadly, how these tricks constitute a novel form of resistance to labor surveillance in the FCs.

\subsection{Workers’ Understanding of Labor-Tracking System and Metrics}

\subsubsection{Figuring Out the Unspoken Systems and Metrics Used by Managers}


The labor-tracking system in the FCs is known to all of our worker participants. Specifically, FC workers (P3, P8, P9, and P14 in particular) know how the tracking system is integrated into their workstations and how it records their work productivity. P3 has learned that the rationale behind logging into the system at the start of the day is to enable it to monitor labor. Although the login system \textit{“might look a little bit different”} depending on the process path bin or worker’s position, \textit{“generally, it [the labor tracking system] starts with you log[ging] into a computer, of course, to keep track of the work that you do. So you scan your badge into a station and, you start your work”} (P3). Participants are also aware of what the labor tracking system tracks and how rates are calculated using the collected data, as P8 noted: \textit{“the labor tracking itself is it can tell whatever you're doing at any given moment. It can tell what item you've scanned, what shelf you've scanned, how long you're going between scanning items. So it calculates a rate”} (P8).

However, not all workers were informed about the labor-tracking system on their first day of work. They had to initially guess how their productivity was being recorded. For some, this knowledge was acquired by accident or during a conversation with a manager or other workers. For example, P9 said,\textit{“when I worked on my first week,..they [management] would tell me my rate. They would say, `Oh, you're doing this amount of packages,' right? So then I kind of questioned it. I was like, `How do they know how many packages I've been doing?'.''} To better understand how the systems and metrics in the FC worked, participants would often search the online communities or Google hoping~\textit{“that would answer my question instead of asking a manager for something”}(P9). P5 describes this experience, \textit{“Not always [Amazon tells workers about the labor tracking system] because there's some workers I know [who] didn't know until me or someone else told them. So sometimes Amazon is not really upfront about it''} (P5). 


Given the lack of transparency on Amazon's part, it was common for FC workers to assume that all labor was being automatically monitored by machines. P6 recounted how she arrived at this presumption: ~\textit{“I don't remember if it was day one, but pretty early on, I learned about ``time-off tasks'' (TOT). They [the management] scanned my badge, and they just told me, `It's to make sure you're not off-task, to make sure the computer doesn't mark you as off-task.' So I just assumed the computer tracked that.''} Upon transitioning from the role of a “picker”\footnote{A picker is the individual responsible for identifying and retrieving items listed on an order form or stored on shelves.} to that of a “tote runner,”\footnote{The role of a tote runner is to bring empty totes to pickers.} P6 realized that there are \textit{“some things that can’t be tracked by the computer.”} After spending so much time speculating about how her work had been tracked, she eventually approached her manager for clarification:~\textit{``When they [the management/HR] started putting me on tote running for pack, I noticed that they [managers] weren't scanning my badge, so I just asked about it. My manager at the time said, `We put it in the computer manually'.''}

\subsubsection{Differential Tracking based on the Position of a Worker: Manual versus Automatic Tracking}

As P6 mentioned in the previous section, not all work can be tracked by the system. This limitation means that certain work activities are not on the labor-tracking system’s radar and, correspondingly, are beyond the scrutiny of the productivity rate system: \textit{“So some of the stuff, we don't have labor tracking metrics for and we don't have systems in place to where it tells you whatever you're scanning. We have batching, which essentially, like I said, is preparing the orders.…That is what we call an `indirect labor' task because it's not directly tracked”} (P8). In contrast to the concept of ``indirect labor,'' ``direct labor'' or ``direct role'' connotes work tasks that entail direct handling of products, such as stowing, picking, or packing. P4 explains that direct labor is subject to extensive monitoring by tracking systems: \textit{“Only your direct value adding processes are heavily tracked as rated productivity. Being receive, stow, pick and pack, that's it.”} 






Direct labor or direct roles, then, are the only labor activities that a computer can understand; indirect labor is not computationally intelligible (P1, P6, P7, P8, P13). This distinction is particularly notable to FC workers as P6 recounts: ~\textit{“A direct role like a packer, their work is being tracked by the computer because it's something that can be quantified and something the computer has to track anyways.…And then indirect roles are things like restocking stations or clearing empty totes. That cannot be tracked by the computer. Because we don't use a computer to do our jobs.''} P13, who has an indirect role, told us that she manually tracks her labor by putting a required labor code into the tracking system:~\textit{“As far as the computer can tell, looking at my time I'm doing nothing. I have to manually tell it, Hey, I'm teaching people, there isn't going to be a lot going into the system from me. So that explains any time that looks like you're off task because you're not doing something that is tracked.”}



\subsubsection{Numerous Technologies Integrated to Coordinate Mechanisms for Tracking Labor}



As the descriptions in section 4.1.2 show, labor monitoring in FCs is intricately linked to the technologies that workers use. As mentioned, the labor-tracking system is a technological assemblage comprising multiple devices, most notably the A-to-Z app and scanners that workers are required to carry at all times. The separate elements of the labor-tracking system are interconnected in such a way that they can both facilitate and monitor workers' activities (P9).

Among the devices mobilized in the FC, scanners are essential for most of the labor force (P1, P4, P3, P8). P4 describes how scanners are pervasive in the FC:~\textit{“So all the processes either had a wireless scanner, a mobile one for in the bins, and the roles at the stations, like working with the conveyor belts for pack or receive, they had a scanner attached to a screen which had the same exact information the hand scanner does.''} Every item, shelf, tote and workstation has its own barcode; in fact, every worker does as well: ~\textit{“we [workers] have to scan barcodes on everything all the time''} (P1). Scanners are the backbone of an FC because they collect and process barcode data in different places in a particular order:~\textit{“There's barcodes on the boxes. And then there's, of course, barcodes on the bins and of course, on the equipment you use and on the different pods that you might stow into. It's basically you scan a barcode on a box, and then you scan the pod that you're going to stow it into, and rinse and repeat, pretty much''} (P3).

As P4 noted above, each worker’s badge has a barcode, which workers are asked to scan at the beginning of their shift. P8 describes how scanning one's badge enables the system to track their activities:~\textit{“Once you're at the scanner, you have to log in to your Amazon employee account, which is you just log in with scanning your badge as the password, because there's a little barcode on the back of the badge. So after that,…you'll go to the labor tracking screen.”} Badge barcodes also allow workers to access certain apps and devices based on their role, which, at the same time, creates another opportunity to be tracked by the system. Several workers (P1, P2, P6, P7) told us that their badges are like their identity:~\textit{“So your badge is everything. You don't want to lose your Amazon badge because then you can't access any of the systems. You can't log onto anything. But that's also how they track you, so when you log onto the system and start performing a function. The system automatically recognizes you and what you're doing”} (P1).



Taken together, the barcodes on worker badges, item barcodes, and pod barcodes orchestrate an intricate surveillance mechanism. By scanning badges, managers can check workers’ progress relative to their assigned tasks at any time. P2 describes when her manager randomly showed up and asked about her badge:~\textit{“every now and then someone will come and say, `Hey, may I scan your badge?' Sure,...And they'll scan it and be like, `Oh, okay cool. You're Waterspider today. And I know what you're doing right now, cool…' If you're not there, you're in trouble.''}

Similarly, workers responsible for handling items can be monitored by tracking the precise location and status of each item through an internal application. P7 explains how “the Pack app” concurrently monitors both items and workers:
~\textit{“you can scan an item, you can scan a tote, a location, and it will show you the history of that location, who interacted with it, who put things in, who took things out, what's supposed to be in this tote.”} The labor-tracking system uses these data to calculate a productivity rate for each worker. P2 elaborates how this metric, which is known colloquially as “scan to scan”, works in practice:
\begin{quote}
“So it's called scan to scan. Between each scan, it monitors how long it takes you to scan the next item. So if you say, `I have a package, it says I have 12 items to put in there.' Well, it should take no more than five seconds to construct, or half construct the box. In the end once you get the box, it should take no more than three to five seconds.”
\end{quote}

An individual worker's productivity rate serves as a reminder of the expected pace of their work, a data point that a worker is constantly reminded of throughout their shift. A key outcome of all the embedded scanners, barcodes, digital data, and productivity metrics in the FC is the contention by many workers (P1, P7, P12) that their physical work is actually primarily virtual. This sentiment is reinforced by a work regimen that requires logging into the system at the start of each work session, and is further compounded by the fact that Amazon considers its \textit{“biggest issue”} (P12) to be a mismatch between the internal inventory and the online store. P1 expresses how working in the physical world with a scanner and numbers is like working in a virtual world:

\begin{quote}
“Because every action within Amazon has a reaction virtually within their system. Even all those people in direct roles who are just packing packages, they are doing something in the system virtually, …and I think that's a big part of what Amazon does in their work and fulfillment centers. It's like yes, we're doing all this work physically, but what is it being within the numbers and data? Because it's all a lot of integrated technology”.
\end{quote}


\subsection{Workers' Perspectives on the Numerical Data Generated by Systems and Metrics}

In this section, we show how the numerical data produced by the labor-tracking system's metrics are perceived by workers, and how any discovered errors or discrepancies require human intervention to resolve.

\subsubsection{System Malfunctions and Unjust Metrics: Discrepancies between the Reality and the Data}

Nearly all workers note that the labor-tracking system often fails to accurately reflect workers' activities and inputs because it frequently malfunctions. For example, sometimes workers are logged out of their scanners suddenly (P5):~\textit{“After a certain amount of time you might have to re-log into your scanners. So, that causes a little bit of a problem because you might have to pick something on your way..Then you find out while you're doing that, oh, it logged me out.”} These kinds of malfunctions are a major concern for workers because system misrepresentations in records such as “Time off Tasks” (TOT) can lead to a warning or penalty (e.g., P9). The most problematic misrepresentation issue for workers, however, is the inaccurate logging of their productivity rates:~\textit{“so it tracks if we missed that time or not, it will count how many misses that we got. So if technology, like the conveyor lines and stuff go down and our productivity is halted, that's bad news for me because then I'm going to have more misses because the facility couldn't continue to process the customer shipments”} (P1).

Beyond inaccuracies, workers also criticize tracking metrics as being intrinsically flawed. For example, P13 suggests that when the same metrics are applied to all tasks they fail to differentiate between the variable intensity of different tasks:~\textit{“I would say reward people for doing those large items that are hard to process, reward people for handling Problem Solve items that they can't actually fully complete because there's something wrong with it [the item]…but I feel like Amazon's current implementation of the rate tracking is surprisingly minimal for how complicated everything an Amazon is.”}  Largely, workers feel that the current tracking system is broadly unfair. P9 expresses frustration with the fact that FC metrics are not aligned with the conditions of specific working contexts (e.g., workers switching tasks), nor can they accommodate unexpected situations (e.g., emergencies, weather disasters). Beyond this, he is upset that tracking metrics have the power to compound a small mistake (e.g., being one minute late) into a disproportionate penalization (e.g., deduction of one full working hour):~\textit{“sometimes I either clock out a bit too late, like one minute too late, and it adds up, they say negative one hour.”} Although the metrics often do not make sense to workers, they nevertheless appear to abide by their logic, as can be seen when workers speed up their work to recover productivity rates that were slowed down by their break times:
\begin{quote}
“So because our break times are taken into account for our rate, so for that 15 minutes that you're in the break room you have to make that up at some point during the day. You're not excused for that 15 minutes from the rate, which is a little strange. But for some people the trade off of having to work a little bit harder for a few minutes is worth it” (P13). 
\end{quote}

\subsubsection{Human Involvement: Interpreting Discrepancies and Manipulating Data}

As noted, technical errors and unjust metrics often lead to discrepancies between tracked data and actual worker activity. To resolve these discrepancies, workers typically need to appeal to their managers, who  monitor and control these data closely. Correcting any erroneous numbers or records requires workers to be able to account for their activity. For example, when the tracking system detects more than 3 to 5 minutes of work inactivity (i.e., Time off Tasks (TOT)), the worker has to explain to their manager exactly what happened so that they can prove that the system logged a TOT by mistake. P4 calls this “bridging dead time”:~\textit{“So that's what I talked about earlier with the time off task, right. So between scans retract as dead time, you have to bridge that dead time. Bridge being basically explained what the problem was. Managers will use that understanding to, one, absolve the associate of any misdoing so they know, 'Okay, this is actually not your fault. That's fine.'''} Otherwise, if more than 2 hours of TOT are recorded, this error would bring on an \textit{“instant termination”} (P4) for a worker.

P4’s quote alludes to the fact that managers typically make the call regarding data veracity. It is up to them to assess whether the system-recorded TOT is justifiable or whether a worker's TOT record can be cleared:~\textit{“the managers will want answers from you for what were you doing during these 20 minutes of inactivity, and you'll have to explain yourself and that can have varying levels of success. Sometimes they'll believe you and just leave it alone. Sometimes they'll be like, `Oh, well, that's your fault.' Even though it's not my fault. They'll give you a verbal warning or whatever, and they can escalate that if they want to”} (P12). In the FC, workers’ labor data is largely at the mercy of managers’ discretion. As such, some workers (P2, P8) mentioned that maintaining good relationships with managers is an important factor in assuring that their labor data and records are accurate. For example, P8 talked about an instance when the labor-tracking system mistakenly tracked her working time as a break,~\textit{“the management [was] like, `Hey, why are you on break for an hour?' I'm like, `I wasn't.' Well, I'm lucky because I have a decent relationship with management, so they know. But for other people, it can cause issues.”} 

In addition to clearing and correcting data errors, managers also manually track workers’ labor occasionally–that is, managers are able to create new numbers and records based on their assessments of a worker’s productivity:~\textit{“They [some FCs] have the managers do it [tracking labor] all for them. They [the managers] will get on the laptops and they have access to the page where they can put in your log and then whatever you need to be tracked into”} (P8). In interviews, workers expressed concerns about managers’ judgment of data and numbers, and intimated that managers sometimes exploit the tracking system:
 \begin{quote}
 “[S]ometimes certain managers abuse it just to track you and see how long you're taking on things and constantly want to see if you're moving fast. So, I think that's where it's a problem. I don't think that's necessarily a technology thing. I think it's just sometimes it could be a manager or something that's giving someone a hard time or something like that'' (P5).
\end{quote}
 Managers were thought to manipulate data in the system for their own advantage because they \textit{“can track and code people’s labor activities and their hours super easily”} (P1). P1 continues,~\textit{“And there's a bunch of different options. So what people will do is they will go in and go to specific employees' activity details, go to their little time bar or taskbar, and they can click it and do edit functions.”} In this way, managers often~\textit{“hide hours in the system so that they don't have that negative being reflected on them, that negative TPH (Throughout per Hour, measuring total volume/ total hours), because it's just not seen as, I guess, like great management”} (P1).

\subsubsection{Workers’ Experiences of Numbers: Being Treated as a Number}


Understandably, the pervasive ubiquity of the labor tracking system, barcodes, and productivity metrics in the FC fosters a logic of quantification when it comes to assessing workers’ labor and activities (e.g., breaks, clock-in/clock-out times). This subsection centers on the prevalence of managers and colleagues depersonalizing employees into mere numerical figures. We found that workers’ grievances about being treated as numbers were also widely shared in online communities. According to workers, Amazon is well aware of this kind of treatment and has gone so far as to initiate regular surveys for workers when they log in at the beginning of work. Survey questions tend to focus on how the workers are feeling in the FC, but one question explicitly inquires about the managerial tendency of treating workers like numbers: \textit{“Do you feel like your manager treats you more like a number than a person?”} (P1, P7). The inclusion of this question would imply that this experience is nearly universal among all FC workers. 

Upon reflection, workers acknowledge that the experience of “being treated as a number” is inevitable in a working environment that is obsessively focused on achieving maximum productivity. P13 describes how the company's ever-increasing productivity expectations haunt him all the time:
 \begin{quote}
“At [a specific FC name anonymized] you always felt like your numbers were looming over your shoulder. If you were just barely making rate you felt it and you felt this pressure to work yourself harder to make up for it. Because even 99.6\% of the rate was still not enough to be considered satisfactory. If you made 99.6 and your manager didn't think you had an excuse for it, they would consider it underperforming” (P13).
\end{quote}
In a similar vein, P1 explains how her mind is preoccupied with numbers on a screen during work, even though the numbers are not always accurate:~\textit{“Up on the screen, it shows you down in the corner, it says UPH (Units Per Hour) and this number, and it's constantly fluctuating...I don't like that because it just puts too much pressure on a number in my mind, for it to be right there for me to be able to look at it all day long,…It would go up and down constantly, and it's not the most accurate number either.”}

Despite the obvious pressure and alienation that comes from being treated this way, some workers say they can sympathize with the managers who are numerically oriented. This sympathy likely stems from the experience of assisting managers who are responsible for overseeing and inspecting various kinds of metrics computed by the labor-tracking system. P4 describes this experience: ~\textit{“whenever you make a move to the virtual inventory. Hopefully it corresponds to the physical inventory, but that's all logged, the associate, timestamp, where it was, when it was, what it was, what the action was…We can understand that people who pick out of this aisle are 10\% slower”}. He continues describing the way in which dealing with so many numbers naturally compelled him not to ~\textit{“care who it was [picking in that aisle].”} He suggests that this desensitization led him to become oblivious to the workers in the physical world:~\textit{“It's like an abstraction of who the person is.”}
  

P7 shared a similar account when she was assisting in the Pick process:~\textit{“I think I've fallen into this [treating others as a number]  too. I realize as well, but when you're working as a process assistant, you have to stop trying to think of everyone as a person, because you're just trying to calculate how much rate I can make if I can get through this day and stuff.”} Although she keeps reminding herself not to treat her fellow workers as numbers, she gets frustrated by how easy it is to forget:
 \begin{quote}
“whenever I had someone working for me as a Picker, I was very concerned [about if she forgets about that again]...I was trying to balance out their days so I'm not having them pick all day…But I know that there are some days where I forget about that completely where I have such high volume and then I just send someone picking for 10 hours a day and I forget to even say hi to them. I feel like when you're doing that, that's when you start to feel more like a number than a person.”
\end{quote}
As implied here, heavy workloads and impossibly high productivity expectations also drive people to get stuck on numbers.  

Understandably, what many workers want most from the company is simply to be treated as a person. P3 believes that this goal cannot be achieved without management making an effort:~\textit{“I guess the first thing I could think of is probably... be to make sure that the workers are treated like people and not just numbers, and that managers and human resources (HR) can be understanding of that''} (P3). Similarly, P13 posits that ~\textit{“the best way”} to empower workers is for the company to ~\textit{“tak[e] steps towards making us feel less like being just another number.”}


\subsection{Workers’ Resistance to Tracking Systems and Metrics: ``Tricks'' Shared by Workers}

The emergence of a dehumanized quantification logic that reifies workers as numbers is not surprising in a work context strongly defined by the use of labor-tracking technologies in the service ofoptimizing worker productivity.  In this section, we focus on how workers react to this quantification logic as well as to the other organizational regulations aimed at increasing their productivity. Our analysis reveals that workers employed “tricks” to help them “artificially” increase their productivity rates or extend their break times. These tricks, which might otherwise be considered widely shared hacks, show that FC workers actively deceive the labor-tracking system and re-purpose warehouse items, equipment, and spaces to gain advantage and benefit themselves. 

\subsubsection{Increasing their Productivity Rates ``Artificially''}


The most frequent tricks by FC workers in our data center on ``cheating'' the productivity rate and TOT (time off task) systems. The goal here is for a worker to boost their rate or optimize the results from other productivity measurements, such as UPH (units per hour) or TAKT (average time to scan and stow an item). Executing the trick involves deceiving the task assignment system via ``cherry-picking'' tasks. Cherry-picking messes with the system, which is programmed to distribute tasks equitably across all workers. P13 describes the trick in more detail:
\begin{quote}
“Being with Amazon for as long as I have, I have heard of a lot of various different tricks. The most common ones would be ones that people use to artificially keep their rate up. So at [a specific FC name anonymized]  where I was picking, people typically call doing these sort of things cherry picking. But the way you would trick the system at [anonymized FC] to get better rates would be if it put you on a pick path that was just too spread out, you weren't getting your items fast enough in, people would log out of the scanner entirely. They would wait five minutes and the system would be seeing that this path was going unpicked and that no one was assigned to it, and see that those orders have to go out and it would assign it to someone else. And after that five minutes you'd log back in and it would reassign you a different path, and if yours was bad you'd probably get a better one.'' 
\end{quote}

Another trick is called “virtual picking.” This is where workers take advantage of the “scan to scan” mechanism---the time measured between scans---to increase their rate. Instead of moving around the warehouse to find and scan the next barcode, virtual picking involves printing all of the barcodes again and scanning them in a single sitting:
\begin{quote}
``I use a lot of tricks because I'm concerned about my rate…when you're picking an item you have that process where it [the picking system] tells you to scan a tote, scan the location of the item, and then you scan the item. Because I have a laptop cart, all the laptop carts come with a printer, so I will print out the barcode for the location, the barcode for the item. So I don't have to physically stand where the item is and scan the location there…. I can just stand there and do that [scanning the printed barcodes] instead of physically going to the area. …So then because of that, I get to boost my pick rate really high if I'm just standing there going really fast too'' (P7).
\end{quote}

As alluded to in P7’s explanation, tricks enable workers to \textit{“find [your] own way to do the work rather than following what machines instruct.”} P7 says that without tricks like virtual picking,~\textit{“then they [pickers] have to keep making the same trips back and forth just because that's what the computer tells them to do.”}

Tricks can also help workers improve their rate by changing the order of their work. P12 proudly explains a trick that dramatically lowers his TAKT time---the average time to scan and stow an item:~\textit{“One trick that we do to make it look like we're stowing really fast is that, like say if I have 10 books that are the exact same book, instead of scanning one, putting in the shelf and then scanning the shelf and then repeating that 10 times, I'll grab one book. I'll scan the barcode, scan the shelf, and then I'll do that 10 times in a row.…Then, I'll just put them all in at once as one big pile. As long as you've counted correctly, then the system thinks that you're just stowing really fast and your TAKT time goes down.”} As P12 describes, if a worker modifies the sequence of a scanning process, the system will fail to detect the change and instead account for it as a task executed at an accelerated pace. The result of this error is a favorable reduction in the worker's TAKT.

Workers use the so-called “pick and choose” trick when handling differently sized items. This trick involves choosing what to pick and process first and what to pick and process later so as to improve one's overall pick rate. For example, picking a large quantity of chocolate bars initially would \textit{“really boost up the pick rate,”} (P7) leaving a bit more time to pick big heavy items (e.g., popcorn machines) later:~\textit{“So a lot of our role is just trying to balance the rate by picking and choosing what to process…That's one of our big tricks, the fact that we can pick and choose like that”} (P7).  


\subsubsection{Securing Marginally More Break Time and Space}

Workers also use tricks to take extra breaks or slightly extend their break times. These activities often entail capitalizing on the affordances and flaws of the labor-tracking system. For example, every workstation has a monitor that instructs workers on their tasks; this monitor also has a menu screen that displays the worker's productivity rate in real time. Noticing that the time on display does not influence how their rates are calculated over time led to the trick of using the menu screen whenever they needed to take a short break without sacrificing their rates. P6 explains:~\textit{“For picking, if there's a menu on screen, at least on the picker side, the menu on screen won't count towards the time since last scan…people bring up that screen when they go to the bathroom or something.”}  


Another way of extending the duration of a break involves a clever exploitation of how the system identifies and logs intervals of rest. For instance, P13 utilizes the ``scan to scan'' mechanism to indulge in a “slightly longer break”:

\begin{quote}
``Break times are typically scan to scan, so what people will do is they will have an item in T-[section] and have an item in G-[section] and get to the last screen on both of them where it just wants you to scan the item one more time. When break starts, they'll scan the item in T-[section] and then they'll leave the one in G-[section] unscanned until they come back so that they get a slightly longer break. Because since it's scan to scan, they're getting a bit more time out of it because they're not taking some of that time out to process the next item.''
\end{quote}

Workers can use the allotted time window before the labor-tracking system marks a TOT (i.e., 2 to 3 minutes) for their own benefit, such as taking a bathroom break or making time to consume a snack. When workers collaborate, this trick can become even more powerful: 
\begin{quote}
``If I'm stowing beside my friend, like one day,...I will tell my friend, `Hey, in two minutes or three minutes, can you go to my station and stow an item for me so that the system doesn't track me as time off task?' We'll do stuff like that for each other...because they're literally the station right beside you usually or two stations down. It only takes them 30 seconds to walk over, grab your scanner and stow an item and walk back. In that time they won't count as time off tasks. Because you need to be away for at least two or three minutes to start raising red flags in the Amazon system'' (P12).
\end{quote}

Another trick that exploits a loophole in the system involves the new clock-in and clock-out system, called the A-to-Z application. This system was instituted following the outbreak of the COVID-19 pandemic and enables employees to remotely punch in and out of their shifts using their smartphones. P9 shares how people have learned to exploit the A-to-Z app in recent years:~\textit{``Anyway, their system with clocking in is pretty flawed. During COVID they had this system where you could clock in using geolocation. It was good in the beginning, but I knew there were people that would clock in using a fake GPS app. And they would pretty much come in late or leave early.''} 

Not all tricks are effective nor are all able to consistently evade detection by managers. Some well-known tactics, such as taking refuge in the restroom during work hours and hiding headphones under beanies, are now likely to be spotted:~\textit{“there's tons of tips and tricks that people try to use to cheat the system in Amazon. I've had friends who were working in the problem-solving department and stuff who got fired immediately because they got caught sitting in the break room while they were clocked in”} (P1). Moreover, workers in indirect roles---roles with tasks that cannot be tracked by the system---cannot rely on system exploitation tricks but rather have had to develop tricks in direct relation to their managers.


Workers develop non-technological tricks too. In particular, they actively repurpose workstation spaces, equipment, and supplies such as boxes and totes to create their own breakrooms. P5 shared that workers resort to using boxes or containers as a means of taking a brief respite because there are no chairs in their workstations and the breakroom is in a remote location. P9 tells how heavy packages can transform into viable seats:~\textit{``We constantly get these cases of water so we pretty much use those as seats sometimes to just sit down in a way to recover our legs. But other than that, if you get no sturdy, heavy packages or whatever, you just had to sit on the floor.''}




\section{Discussion}

This section examines FC workers' experiences under algorithmic management in two parts: first, we explore the interrelationship between managers as “context probers” and workers’ reliance on online communities for algorithmic sensemaking (Section 5.1). Then we apply the concept of work games to show that workers' development of “tricks” can be viewed as forms of consenting and non-consenting resistance (Section 5.2). Together, our discussion highlights new dynamics in the interplay among algorithmic control, managerial intervention, and worker agency.

\subsection{Characterizing Warehouse Workers' Experiences with Algorithmic Management}

The first point of discussion evinced by our findings is their relation to previous studies of algorithmic management. Our FC data reveal distinctive algorithmic patterns, both in contrast to the research on platform-mediated work as well as work that occurs in more traditional knowledge work environments. Analytically, these distinctions suggest that algorithmic management is not a uniform phenomenon, but rather should be understood as uniquely configured to an organization's work culture and the specific implementation of algorithmic tools in the work environment. 

\subsubsection{The Manager as a ``Context Prober'' and Source of Unpredictability}

The first distinction regarding algorithmic management in an FC involves the role of the manager. Previous literature has documented a shift in managerial roles and worker interactions when algorithmic management tools are introduced into the work context. For instance, Craig Gent’s research~\cite{gent2018politics} highlights managerial distantiation (i.e., the absence of managers from the shop floor) in response to the adoption of algorithmic management techniques. Gent also underscores the pervasiveness of algorithmic management across all organizational levels—managers and workers alike are alienated by the underlying logic of these systems. The increased disconnection among workers ends up ceding decision making authority to the algorithms, destabilizing managerial authority as a result. In our study, by contrast, we see something other than the diminishment of middle managerial authority~\cite{lee2015working}. We unpack and illustrate our detailed understanding of the shifting managerial roles in algorithmically managed FCs below, drawing primarily on the perspectives of managed workers.

Like many prior platform studies~\cite{shapiro2018between,ajunwa2017limitless,bernhardt2023data}, our data confirm that FC workers are constantly being monitored and evaluated by algorithmic systems. However, the ubiquitous interaction with other humans, especially managers, in the FC affects their perceptions of algorithmic evaluation, which in our study is termed ‘productivity rates.’ Indeed, as we have described, FC workers were tracked throughout the work environment, but they were also monitored by their managers-a managerial pattern we refer to as “double monitoring.” In an FC, human managers' oversight (e.g., on workers' attitudes and work ethics) is vital in shaping how work is accomplished because it is a setting ``where algorithmic systems are not the primary means of organizing work''~\cite[p.4]{jarrahi2021algorithmic}.

As we have shown, managers play a large role in addressing algorithmic discrepancies in an FC. Workers cannot merely attribute unsatisfactory outcomes (e.g., an unfair task assignment or an inaccurate job recommendation) to a “mistake” made by the algorithmic system~\cite{lee2015working}; instead, they must make an entreaty to their manager. Sometimes workers report that managers bypass them entirely in an attempt to inspect the working lines, validate workers’ recorded data, or fill any data gaps. The intervening position of the manager in the FC resembles the role of the “algorithmist” proposed by~\citet{gal2020breaking}. Algorithmists are actors who interpret and mediate algorithmic logic and data. In the FC, managers act as quasi-algorithmists when they take on the role of “context prober''---the person responsible for correcting or modifying a worker’s captured data in accordance with ongoing work contexts. Instead of fully interpreting and mediating data, they merely probe it to determine whether any unusually low rates are the result of an aberration.

Assuming the role of a context prober reorients managers'  relationships with their workers. With this power, a manager becomes a crucial aspect of a worker's algorithmically-managed experience as well as his or her ability to maintain a high productivity rate.  While platform gig workers are subject to the “tyranny of the algorithm”~\cite{lehdonvirta2018flexibility} or “algorithmic despotism”~\cite{griesbach2019algorithmic}, FC workers are far more subject to the way that managers handle data discrepancies. FC workers perceive that managers will either amplify or diminish an algorithm’s influence on them, depending on whether the managers strictly measure the context or ignore the gap. Managers end up being the only source of unpredictability for workers. This human unpredictability in the FC runs contrary to the experience of rail-hailing platform workers~\cite{cameron2022making}, for example, who report having greater control over humans than machines. 


The tight coupling between managerial intervention and algorithmic tracking in the FC suggests that workers here experience ``conditional algorithmic management,''~\cite{wood2021:algorithmic} in which managerial agency is co-mingled with algorithmic agency. Prior work  ~\cite{woodLehdonvirta:2021antagonism} finds that eliminating the presence of a human manager can provide workers with a sense of autonomy but also risks increasing their dependence on the platform, resulting in what the researchers refer to as ``subordinated agency''~\cite{woodLehdonvirta:2021antagonism}. In contrast, our study finds that workers perceive themselves to be simultaneously subordinate to both human and algorithmic systems with only an occasional sense of enhanced agency, if at all. Workers' perceptions of losing control over their rates is directly connected to their relationships with managers, because managers' power to intervene in metrics ultimately gives them the power to make workers feel subordinate. This complex view of algorithmic subjugation destabilizes the current understanding of algorithmic management because it contradicts the viewpoints of some logistical media theorists~\cite{rossiter:2016software,beverungen:2021amazon} who posit that warehouse managers operating under algorithmic management would be bereft of agency and decision-making abilities because decisions made by ``disinterested data-crunching algorithms''~\cite[p.125]{rossiter:2016software} are perceived to have higher reliability.


\subsubsection{Relying on Online Communities rather than In-person Interactions with Colleagues}



Our second point for discussion involves work related to skill-building and knowledge sharing. As our findings indicate, FC workers have a solid understanding of how labor-tracking system functions in relation to their handheld devices. Workers know what these tools measure and how the derived metrics are used to calculate productivity-based scores. This recognition falls short of full knowledge, however, because the company and its managers are not always transparent about the system and its range of metrics. As a result, workers are forced to navigate these algorithmic environments on their own. This challenge has been discussed in the literature on platform technologies in recent years ~\cite{kinder2019gig,tandon2022hostile}. In one study platform freelancers describe practices of monitoring the platform’s functions, looking at the client side of the platform, and using social channels to explore and learn how algorithms are affecting their work arrangements. These kinds of practices have been dubbed as “sense-making” strategies~\cite{jarrahi2019algorithmic} and tactics for “guessing the system”~\cite{mohlmann2017hands}. With regard the the latter, Mohlmann and Zalmanson~\cite{mohlmann2017hands} show how rail-hailing platform workers developed stories and myths as a means of addressing the uncertainty and information asymmetry they encountered in relation to the algorithmic system in place in their work. 

Our study returns to this topic to highlight how little attention has previously been paid to the specific mechanisms by which workers come to comprehend the algorithmic management system in which they are embedded. Related empirical studies have mainly centered on explicating the relationships between distinct work practices and algorithmic management rather than addressing individual workers' experiences with algorithmic management as a whole. While there has been some acknowledgment of that workers often experiment to learn system capabilities and limitations, these accounts have typically focused on workers' broader sense of opportunities within the system. In sum, this research leaves the details of workers' sensemaking processes largely unexamined.

In practice, experimenting with systems is typically an individual act, done by workers in their particular situated context. However, our study highlights the role and import of collective sensemaking, in particular within online communities. Online community interactions enable workers to acquire knowledge and insights into the underlying algorithms of the systems and metrics at the FCs, albeit in a somewhat decontextualized manner. This finding largely resonates with many accounts of platform gig workers~\cite{irani2013turkopticon,ma2022brush,salehi2015we} who also report a heavy reliance on online forums as a valuable source of information. ~\citet{kinder2019gig} illustrate that “communal sensemaking” is strongly manifested in their study on platform workers who collectively try to resolve their confusion about how their job success score is calculated. It is noteworthy that while interpersonal interactions at work, such as daily interactions with colleagues or managers, may seem to provide more opportunities for our workers to learn about the functioning of labor-tracking system, online communities remain the primary source of information for them. 

Surprisingly, few studies of traditional workplaces address the formation, reliance on, or even the existence of online communities as a collective forum for knowledge sharing among workers. For example, investigations into environments such as call centers, courtrooms, or newsrooms rarely acknowledge the role or impact of such communities. This omission may be attributed to the scale and distribution of the workforce; workers engaged with expansive algorithm-driven systems—such as those operating across multiple Amazon warehouses in the U.S. and Europe similar to Amazon, or in widespread gig economy platforms like Uber—often find themselves deeply influenced by and dependent on these virtual networks. Such environments, perhaps more common in big tech industries, uniquely foster the reliance on online communities for support and information exchange.

In fact, the FC workers in our study rarely exchange tips or advice with their coworkers in person. Intense work schedules and fast-paced work environments, coupled with the isolating set up of most workstations, result in limited opportunities for co-located, inter-worker interaction at the FC. Online communities, thus, become important spaces for workers to gain awareness of peers' experiences with labor-tracking system as well as a significant mechanism for cultivating ``algorithm sensemaking''~\cite[p.5]{mohlmannn2023algorithm}. Our observation echoes the norm laid out by Gent in his study of food distribution center work~\cite{gent2018politics} where worker communication, particularly verbal interactions, whether direct or indirect, are strongly limited. Both our study and Gent's underscore how algorithmic management can amplify traditional workplace restrictions through technological means. Workers are required to continuously attend to instructions from devices or engage with hand-held devices monitored by management, which effectively isolates them at their workstations surrounded by monitors and machinery. This setup not only perpetuates, but likely intensifies, the pre-existing rule against talking found in conventional warehouse and distribution center workplaces. By feeding data into the algorithmic management system, these devices significantly reduce any potential for social interaction among workers.




Workers’ attempts to gauge algorithms via channels outside of their workplace reveals an underlying trend of low transparency in regard to algorithmic management. Companies often design algorithms to be inscrutable on purpose so as to prevent employees from manipulating the system~\cite{jhaver2018algorithmic,eslami2017careful}. Together with other studies on platform gig workers, our research demonstrates that making algorithmic systems opaque does not deter workers from seeking knowledge, whether on their own or in solidarity with their coworkers.

\subsubsection{Individualized and Self-Organized Reactions}

  
Within the context of platform work, previous CSCW scholarship has documented and classified various strategies and tactics that workers use to resist algorithmic power (as we detailed in Section 2.2). Like the platform workers in these studies~\cite{chen:2018Didi,mason:2019chasing}, we also see FC workers attempting to circumvent (e.g., log in and out of the system to have an easier task assigned to them) and/or manipulate the tracking system (e.g., leaving the last item for the system to process so as to take a slightly longer break). Recent CSCW literature has begun to spotlight platform workers’ collective forms of resistance and manipulation~\cite{curchod2020working,tandon2022hostile,calacci2022bargaining}. For example, Kingsely et al.~\cite{kingsley:2022give} show how YouTube content creators utilize their collective ``voice'' when responding to (de)monetization algorithms. By requesting a response from the platform company, their collective actions directly challenge the company to address the inequitable algorithmic content decisions. In addition, these workers are galvanized through diverse social media platforms such as Twitter, as opposed to being confined to their online forums. Beyond the platform context, one study~\cite{gent2018politics} shows how a small group of workers created ``collective guile''---a shared situational understanding---which enabled them to coordinate work slow-downs. 

By contrast, our study with FC workers found little evidence of collective action---or “algoactivism,” as~\citet{kellogg2020algorithms} refers to it. This lack of organized responses to algorithmic management within the FC context highlights another key distinction in our research. FC workers’ acts of resistance, or “tricks,” are largely characterized by self-organization and individualization, notwithstanding the sharing of existing tricks among other workers in their online communities as mentioned previously. 
Labor-tracking systems' intensive quantification of individual worker's productivity may contribute to reason that many of these tricks are individual-centered games. It could also be possible that the lack of collective action is due to FC workers’ unique circumstances, in which any attempt to form a union or organize strikes is strictly monitored and thwarted by the company~\cite{massimo2020struggle}. In the following section, the implications of our workers' various resistance tricks will be discussed and unpacked through the lens of the work games.




\subsection{Work Games and Workers’ Determination to Win}

Research on workers' forms of resistance in the face of algorithmic management has been well documented in the CSCW literature, and yet we believe that it falls short in explaining why workers resist and what causes this resistance to persist. To gain greater analytical perspective, we look outside of our field to identify parallel discourses occurring within sociology, labor, and organizational studies. These discussions have, to date, focused more explicitly on the interrelation of worker resistance and various forms of workplace control. In reviewing this literature we find a strong resonance with the concept of `work games' (e.g.,~\cite{ranganathan2020numbers,sallaz2015permanent}), which we apply here to help advance the conversation about why FC workers and others appear to actively engage in `their own exploitation'~\cite[p.xi]{burawoy1979manufacturing}. The idea of gaming work has influenced scholars in various fields, and has been studied in contexts including hotels~\cite{sherman2007class}, casinos~\cite{sallaz2009labor}, service industries~\cite{mears2015working}, and ride-hailing platforms~\cite{cameron2022making}. Collectively, this research reveals that workers often strategically modulate their efforts to achieve specific outcomes. When we apply the concept of work games as an analytical lens to interrogate our distinct study context---which, as already noted, includes algorithm-driven decisions on discipline and termination as well as monitoring workers' physical exertion under direct human supervision---we can see added nuance in the ways that workers engage with their tasks. For the record, modern FCs are markedly different contexts from the factory floors or warehouses of the 1950s, the context in which the original idea of work games derives, as well as from contemporary forms of platform-mediated work. Yet, we believe that this framing is nonetheless salient for its capability to offer new insights on worker resistance and workers' subjective experiences under algorithmic management.


\subsubsection{Work Games as a Lens for Understanding Worker Behavior} 

The field of industrial sociology has long recognized that the context of work is laden with power dynamics. Several decades ago, researchers showed how workers often ``gamed'' their environments to make time pass quickly and to endure arduous and meaningless tasks. For example, Donald Roy identified a game called “banana time”~\cite{roy1959banana}, in which drill press operators at a garment factory collectively dealt with their monotonous work. The game is quite simple: each worker started their break time by stealing a banana from a colleague’s lunch box and calling out “banana time!” 

Later, sociologist Michael Burawoy~\cite{burawoy1979manufacturing} formally articulated the concept of “work games” based on his ethnographic study with factory workers. During his observations, he recognized that workers were engaging in a worker-initiated piece rate game called “making out.” In playing the game, workers attempted to reach beyond the quota level to maximize their earnings by improvising tools and speeding up machinery as they received bonus pay calculated by the piece-rate remuneration system. Burawoy's key insight was an examination of how and why workers consent to extracting surplus value of their labor. To this end, he found that the making out game provided workers with social and psychological rewards, most notably countering the frustration that arose from the repetitive nature of their work. He described the workers, including himself, as engrossed in a game that they desperately sought to “win”; making out became a ludic means of obtaining social and emotional rewards on the job. Burawoy notes, \textit{“Once I knew I had a chance to make out, the rewards of participating in the game...absorbed my attention, and I found myself spontaneously cooperating with management in the production of greater surplus value”}~\cite[p.64]{burawoy1979manufacturing}. Becoming absorbed in winning a game serves to divert workers' attention away from their exploitation. Burawoy went on to confess, \textit{“I was struck by my own absorption into the game that I knew to be furthering my exploitation. I was not coerced into hard work. As my day man told me on my first shift, ‘no one pushes you around here’ and he was right. Nor could the extra money explain my devotion to hard work.''}~\cite[p.197-198]{burawoy1979manufacturing}. Building upon Gramsci’s idea of hegemonic workplace—where coercion is ingeniously comprised of both persuasion and consent—Burawoy argued that \textit{“consent rather than fear ruled the shop floor”}~\cite{burawoy:2012manufacturing}. A work game, as such, simultaneously enables workers to \textit{secure} as well as to \textit{obscure} their exploitation. 

We draw on two unique aspects of Burawoy’s concept of the work games to describe the activities of the FC workers in our study. First, a work game allows us to view a worker’s consent as inherently created within the context of the game. Second, despite their affordances for workers, work games are actually closely aligned with managerial goals. Any game has rules and playing a game requires adherence to its rules. As Buroway puts it, “the very act of playing the game simultaneously produce[s] consent to its rules''~\cite[p.193-194]{burawoy1979manufacturing}. He further suggests that anyone who decides to play a game cannot question its rules and goals once engaged. Because a work game is built around the rules and structures at a workplace, Burawoy sees that playing a work game inevitably entails workers’ consent to their work. Workplace games are, in his view, worker-initiated rather than imposed from the top-down by the management: workers exercise their agency in establishing and playing the game. Burawoy argues that small winnings for workers disguise worker’s exploitation and managerial controls over them. 


Through the lens of work games, we can characterize the tricks that FC workers play in three ways. First, within the context of a game, workers become agential actors in their own narrative. Within their game, they can claim a certain degree of autonomy and control over the work process, even as this ultimately leads to reinforcing managerial goals. In the context of our study, work games are created and initiated voluntarily by workers; there is little to no managerial involvement in these games. In this setup, Burawoy posits that a worker can be regarded as an ``active subject''~\cite{thompson:1999changing} within the context of workplace controls. From this perspective, a distinction is made between games and gamification, wherein gamification pertains to the utilization of game elements that are developed and incorporated by companies to elicit workers’ active engagement in achieving the company's productivity aims. 

Work games help to reveal FC worker resistance in a second way as well. According to Burawoy, ``the games'' refer to the ways in which workers manipulate a production process for personal gain. Importantly, he observed and experienced firsthand that workers were able to convert a tedious and demoralizing job into ``an exciting outlet for workers to exercise their creativity, speed, and skill''~\cite{mason:2019chasing} when engaging in a work game. Despite its limited opportunities, this outlet, in addition to providing them with a sense of relative autonomy and control, serves as a mirror that reflects their own creative efficacy as a worker. This phenomenon is exemplified in our study by workers' use of the term ``tricks.'' Here they are denoting not only their methods of work process manipulation but are also affirming their individual agency and self-expression. Consequently, engaging in ``tricks'' is not merely a power move, but a key mechanism for workers to satisfy their ``desire for self-determination and self-expression''~\cite{mason:2019chasing}. 

Finally, Burawoy found that the experience of winning the games reinforced workers' commitment to playing and ultimately directed them to continue of the path of producing surplus value for their employers. In this study, we observed that workers' commitments and consent to playing tricks varied, as did the degree to which they acknowledged an alignment with management's interests. In the following section, we describe these various responses in greater detail.

\subsubsection{Workers’ Tricks: Three Forms of Consenting Resistance}

Analyzing FC workers' “tricks” through the lens of work games enables us to identify two forms of consenting resistance and one form that is non-consenting: \textit{active consenting-resistance} (consenting to the rules while resisting the algorithms), \textit{passive consenting-resistance} (breaking the rules without being noticed while resisting the algorithms), and \textit{non-consenting resistance} (refusing to participate in the game).

Before delving into this discussion, let us first define what we mean by the term ``consent.'' We adopt Burawoy's conceptualization of consent, namely that ``playing the game simultaneously produces consent to its rules.'' In other words, no player can play while questioning the game's rules and objectives. This view is similar to that of a number of game scholars who believe that play is ``fundamentally consensual'' or ``voluntary for the player''\footnote{Since our workers are playing alone, the meaning of consent is derived in part from Trammell's definition. In the context of this study, ``the played'' refers to the labor-tracking system and its associated metrics.}~\cite[p.43]{trammell:2020torture}. Consent indicates that workers have the option to engage in ``tricks'' or not; workers can voluntarily choose not to employ ``tricks.'' To add additional nuance here we also follow critical HCI scholars' understanding of consent: consent as a ``sociotechnical condition''~\cite[p.570]{chowdhary2023can}. We do so on the assumption that workers' unique interactions with algorithmic systems leads to the generation of specific games. For example, the forms of resistance exhibited by our workers differ from those of journalists~\cite{christin2017algorithms}, legal professionals, and police officers~\cite{brayne2021technologies}. The latter groups tend to engage directly with the impact of algorithms by manipulating variables or even blocking algorithmic systems. However, our workers, who lack access to the internal algorithmic systems they must utilize, attempt to deceive the system. Despite this difference, we would still assert that our workers consent to the games, which they themselves have produced sociotechnically. Even so, it is important to note that their levels of consent may fluctuate. Therefore, consent as used in this discussion involves a ``nuanced set of ongoing interactions''~\cite[p.4]{nguyen2020challenges} for negotiating the extent of consent given at any particular moment.


The majority of FC employees’ tricks fall under the category of “active consenting-resistance” (consenting to the rules while resisting the algorithms). Using control mechanisms and labor-tracking systems to their advantage, we see workers constructing work games aimed at maintaining high productivity rates or avoiding recorded TOTs (time-off-tasks). When workers use their allotted 2-3 minute time threshold to go to the bathroom without accruing a time-off-task they have won this mini game. Workers also use “virtual picking” and “pick and choose” to achieve high rates by exploiting system flaws or “fissures”~\cite{ferrari2021fissures}, such as the scanner’s inability to detect a process change. In this way, our workers construct games that are difficult to detect and, therefore, easy to escape management control---a form of “covert resistance” according to Cameron and Rahman~\cite{cameron2022expanding}. 

The second type of consenting resistance we see is passive, namely when workers break rules without being noticed while resisting the algorithms. Here the game is about avoiding notice by management rather than exploiting a system weakness. Taking a break at workstations, for example, is not permitted in FCs. As we have shown, however, workers regularly sit on the floor or on a sturdy box to take a quick break; doing so while maintaining a high productivity rate for an extended period of time provides an even higher reward for a worker. Lastly, there are workers who refuse to play games. We refer to this form of resistance as ``non-consenting'' because in this practice workers engage in any method to avoid working under algorithm management. One example in our findings is a worker pretending to clock in with fictitious GPS data or hiding in the restroom.

Looking anew at our examples of consenting and non-consenting resistance through the analytical lens of work games helps us see how workers continue to exercise agency within algorithmic environments. Various forms of worker resistance highlighted in prior studies—such as Kusk and Bossen’s “optimization strategies,” Mohlmann and Zalmanson’s “loopholes in the systems,” and Jarrahi and Sutherland’s “manipulation of platform-work algorithms''—have focused on specific, concrete tactics employed by workers to circumvent algorithmic control or maximize earnings. We contrast these examples by shifting the focus from describing specific forms of resistance to examining how these acts, however ludic and resistant, continue to imply a degree of consent to management goals. Investigating the conditions of algorithmic management in the FC helps to highlight the nuanced boundary between “consent” and “resistance” in workers' actions. The following section explores how work games negotiate this subtle balance in greater detail.

\subsubsection{Workers’ Consent and Resistance as a Dialectic} 

As we have seen in the section immediately preceding, many forms of worker response to algorithmic management take the form of “active consenting-resistance.” These practices empirically demonstrate how workers can both consent and resist simultaneously, and, more importantly, how consent and resistance merge and overlap. Organizational scholars have theorized these kinds of dialectic relationships between consent and resistance in previous literature, with ~\citet{jermier1994resistance} writing that “what might be seen, from one perspective, as resistance, might just as easily be viewed as conformity, compliance or indifference, from another”~\cite[p.4]{jermier1994resistance}. ~\citet{collinson1994strategies} asserts that “resistance frequently contains elements of consent and consent often incorporates aspects of resistance” (p.29). Others, such as ~\citet{fineman1996experiencing}, also contend that consent and resistance are not distinct responses. Rather, they can “coexist in the same form of behavior” (p.87). To wit, resistance can be achieved through consent~\cite{ashcraft2005resistance}.

Research on rail-hailing platform workers’ resistance~\cite{cameron2022making} has called for an investigation into why individual workers choose to play a particular work game. In the context of Amazon warehouses, our findings provide a partial answer to this question. Workers are motivated to develop and implement tricks to maintain their work efficiency, increase their work rates, and have longer break times. We argue that investigating whether workers’ resistance to algorithmic management entails their consent to the work games, and, if so, how and to what extent this is the case, can provide us with a more nuanced understanding of workers’ experiences and reactions to algorithms. We must obtain such insights if we are to intervene and implement deployable changes (e.g., new regulations and policies) to create more democratic workplaces.

\section{Conclusion}

While recent CSCW and HCI studies offer a rich understanding of workers’ experiences under
algorithmic management, these insights are predominantly based on the experiences of workers doing platform-mediated gig work. To broaden the empirical base for algorithmic critique, we conducted an ethnographic study of workers’ experiences with algorithmic management and control within the context of an Amazon fulfillment center. In this context, workers are subject to algorithmic technologies of monitoring and control in the form of a assembled labor-tracking system and its corresponding metrics. 

We showed how the labor-tracking system integrates productivity metrics, a myriad of devices, and product and badge barcodes to keep workers within an algorithmic panopticon. Furthermore, we described how humans are required to resolve the discrepancies between the labor tracked by the system and the actual labor input. Our findings also demonstrate that the labor tracking systems in place in FCs tend to obscure human workers by rendering their labor as numbers. Finally, we detailed the various tricks employed by human workers to resist these labor tracking systems and the imposition of becoming enumerated. We contextualize these findings within the existing literature of CSCW and other relevant fields, elucidating how the distinct experiences and reactions of FC workers are influenced both by the characteristics of traditional work environment (i.e., co-present managerial oversight) and platform environments (i.e., nearly ubiquitous digital tracking).

Through the lens of work games, we discussed how the various “tricks” that FC
workers engage in are simultaneously forms of exerting creative agency and complying with managerial goals. We identified three patterns in workers' work games: active consenting-resistance, passive consenting-resistance, and non-consenting resistance. Based on our analysis, we argue that investigating workers’ concurrent resistance and consent helps us understand workers’ experiences and reactions to algorithms with greater clarity and sympathy. Both are necessary if we are to design a better future for workers in any environment. 



\bibliographystyle{ACM-Reference-Format}
\bibliography{FC-worker.bib}

\end{document}